# From High-Entropy Ceramics (HECs) to Compositionally Complex Ceramics (CCCs)

Jian Luo*
University of California, San Diego
La Jolla, California 92093, U.S.A.
*e-mail: jluo@ucsd.edu

This invited talk will review a series of our recent studies on high-entropy ceramics (HECs) and compositionally complex ceramics (CCCs) and discuss the future perspective. Various single-phase equimolar quinary (five-component) HECs, *e.g.*, $MB_2$, $MB$, $M_3B_4$, $MB_4$, and $MB_6$ borides, $MSi_2$ and $M_3Si_5$ silicides, perovskite, fluorite, and pyrochlore oxides, and (intermetallic) aluminides have been fabricated. We further proposed to extend HECs to CCCs to include non-equimolar compositions and further consider short- and long-range orders, which reduce configurational entropies but offer additional dimensions and opportunities to tailor and improve various properties. We also reported the first dual-phase HECs/CCCs. Using compositionally complex fluorite-based oxides (CCFBOs, which can possess fluorite or defect fluorite, pyrochlore, weberite, fergusonite, and bixbyite phases) as the model systems, we have recently discovered long- and short- range orders, composition- and redox-induced order-disorder transitions (ODTs), and ultrahigh-entropy weberite and fergusonite phases in several 10- to 21-component systems.

1. INTRODUCTION

High-entropy alloys (HEAs) and compositionally complex (or complex concentrated) alloys (CCAs) have attracted substantial research interests since the original reports of Yeh *et al.* [1] and Cantor *et al.* [2] in 2004. See a critical review by Miracle and Senkov [3]. In 2015, Rost *et al.* reported an entropy-stabilized oxide (in the rock-salt structure) [4]. In 2016, Gild *et al.* reported the first non-oxide high-entropy ceramics (HECs) fabricated in the dense bulk ceramic form: a series of single-phase high-entropy metal diborides [5]. In 2018, Jiang *et al.* reported high-entropy perovskite oxides, representing the first HECs with two cation sublattices [6]. In 2020, Qin *et al.* reported the first dual-phase HECs and discovered a thermodynamic relation governing the equilibrium compositions of the two phases [7]. Also in 2020, Wright *et al.* proposed to extend HECs to "compositionally complex ceramics (CCCs)" to include non-equimolar compositions and further consider short- and long-range orders, which reduce the configurational entropies of mixing, but offer additional dimensions and opportunities to tailor and improve various properties [8, 9]. Recent developments include the discoveries and studies of several 10- to 21-component ultrahigh-entropy phases, which can also exhibit order-disorder transitions (ODTs) [10-12]. This invited talk will review a series of our recent studies on HECs and CCCs and discuss the future perspective.

2. EXPANDING HIGH-ENTROPY CERAMICS

In 2016, our research group first reported the successful fabrication of eight high-entropy metal diborides [5] (Fig. 1) as a new class of ultra-high temperature ceramics (UHTCs), representing the first high-entropy borides (as well as the second bulk HECs and the first bulk structural HECs). This report [5] inspired numerous follow-up studies of both high-entropy diboride [13] and carbide [13-16] UHTCs worldwide. We have further improved (or developed novel) synthesis and processing methods to fabricate these high-entropy metal diborides [17-19]. Interestingly, we demonstrated that dissolving and stabilizing softer components (such as $WB_2$ and $ErB_2$, which are either unstable or have little solid solubility in normal binary diborides) can enhance the hardness of high-entropy metal diborides [19, 20]. This serves an exemplar that HECs can exhibit unexpected and improved properties.

As shown in Fig. 1, our research group also first fabricated (in the dense bulk ceramic form): high-entropy monoborides (as a new class of superhard materials) [21], $M_3B_4$ borides [22], and tetraborides [23], as well as dense bulk high-entropy hexaborides [24], a high-entropy metal disilicide (the first high-entropy silicide) [25] and quinary equimolar $M_5Si_3$ silicides (with cation ordering so that they are no longer "high entropy") [26], high-entropy perovskite oxides (the first HECs with two cation sublattices) [6], and yttria-stabilized zirconia (YSZ)-like high-entropy fluorite oxides (as new materials for thermal barrier coatings or TBCs) [27]. Furthermore, we made original contributions (beyond quinary equimolar compositions) in exploring compositionally complex fluorite-based oxides (CCFBOs), including disordered fluorite or defect fluorite oxides [8-12, 27-30], ordered pyrochlore oxides [10-12, 27], and the first ultrahigh-entropy weberite and fergusonite phases discovered [12]. Here, we explored beyond the common quinary (or occasionally 4- to 6-) equimolar compositions to investigate both lower and higher entropy configurations, with non-equimolar compositions or long- and short-range ordering, which will be discussed subsequently in §5. We also reported high-entropy aluminides [31] as the first single-phase high-entropy intermetallic compounds to bridge HECs and HEAs.

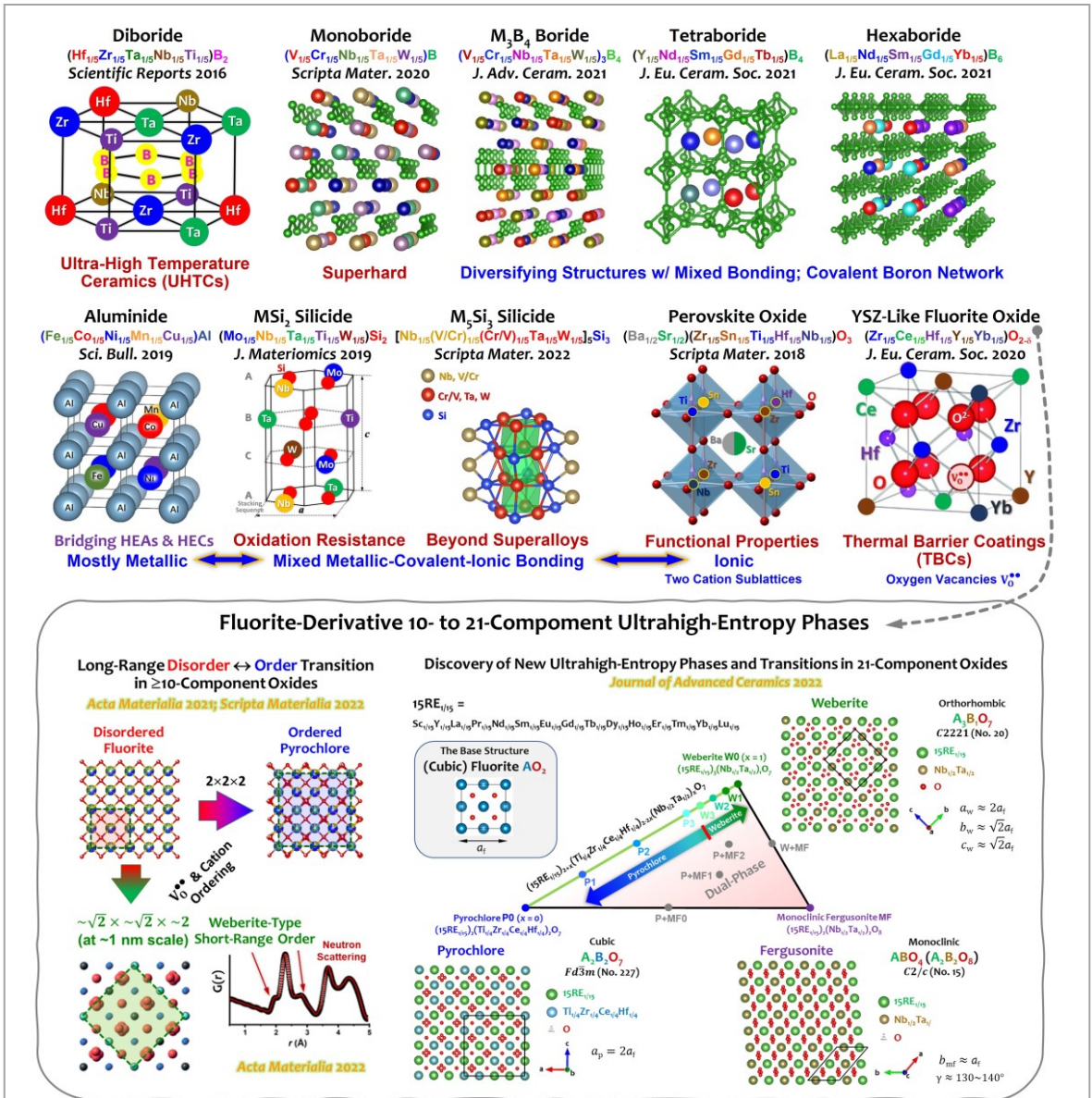

**Fig. 1.** Schematic illustration of a series of high-entropy ceramics (HECs) fabricated in Luo's lab at UCSD with diversifying chemistries, crystal structures, and bonding characters, including quinary equimolar $MB_2$ (a new class of UHTCs) [5], MB (a new class of superhard materials) [21], $M_3B_4$ [22], $MB_4$ [23], and $MB_6$ [24] borides, $MSi_2$ [25] and $M_5Si_3$ [26] silicides, $ABO_3$ perovskite oxides (the first HECs with two cation sublattices) [6], and YSZ-like fluorite oxides (a new type of TBCs) [27]. We also reported high-entropy aluminides [31] as the first single-phase high-entropy intermetallic compounds to bridge HECs and HEAs. Furthermore, we originally explored 5- to 21-compoment compositionally complex fluorite-based oxides (CCFBOs), including disordered fluorite and defect fluorite [8–12, 27–30], ordered pyrochlore [10–12, 27], and newly discovered ultrahigh-entropy weberite and ferguson ite phases [12], which exhibit long- and short-range orders and order-disorder transitions (ODTs).

Here, let us discuss a most recent study of quinary equimolar $M_5Si_3$ silicides [26] as an example, which show some unusual characters (differing from the common disordered quinary equimolar HECs). Here, we fabricated, for the first time, a new type of quinary equimolar silicides, $(V_{1/5}Cr_{1/5}Nb_{1/5}Ta_{1/5}W_{1/5})_5Si_3$ and $(Ti_{1/5}Zr_{1/5}Nb_{1/5}Mo_{1/5}Hf_{1/5})_5Si_3$, both of which are in single-phase homogenous solid solutions. Interestingly, $(V_{1/5}Cr_{1/5}Nb_{1/5}Ta_{1/5}W_{1/5})_5Si_3$ forms the hexagonal γ ($D8_8$) phase, while all its five constituent binary silicides, $V_5Si_3$, $Cr_5Si_3$, $Nb_5Si_3$, $Ta_5Si_3$, and $W_5Si_3$, are stable in the tetragonal α ($D8_l$) or β ($D8_m$) phases. X-ray diffraction, Rietveld refinements, and scanning transmission electron microscopy jointly suggested cation ordering, which reduces the configurational entropy (so that it is no longer "high entropy" by the conventional definition [9]). This work expanded the field of HECs and CCCs by not only discovering a new compositional complex silicide phase but also demonstrating the cation ordering and unusual phase stability.

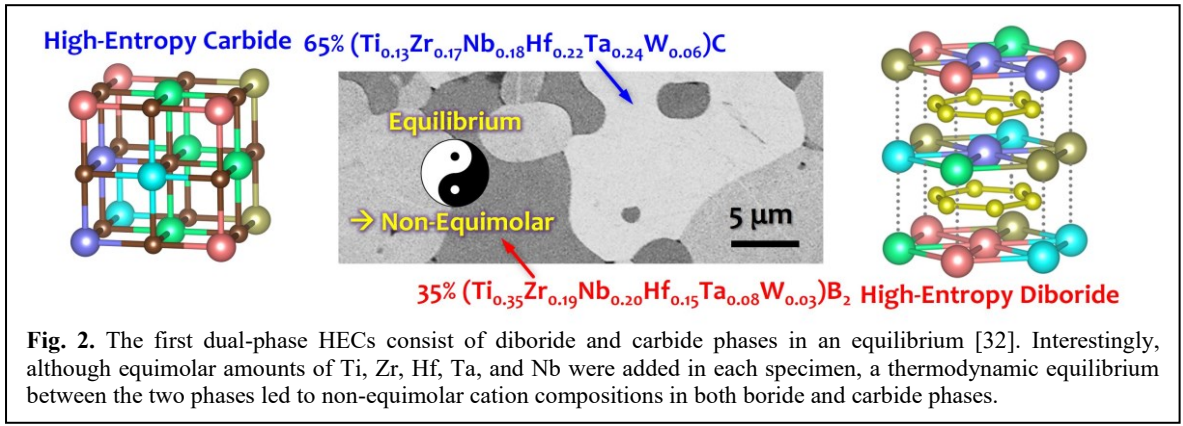

**Fig. 2.** The first dual-phase HECs consist of diboride and carbide phases in an equilibrium [32]. Interestingly, although equimolar amounts of Ti, Zr, Hf, Ta, and Nb were added in each specimen, a thermodynamic equilibrium between the two phases led to non-equimolar cation compositions in both boride and carbide phases.

## 3. DUAL-PHASE HIGH-ENTROPY CERAMICS

In 2020, we reported the first dual-phase HECs consist of metal diboride and carbide phases in an equilibrium one another (Fig. 2) [32]. Interestingly, although equimolar amounts of Ti, Zr, Hf, Ta, and Nb were added in each specimen, a thermodynamic equilibrium between two compositionally complex phases led to non-equimolar cation compositions in both boride and carbide phases (that reduced the configurational entropies).

Here, we discovered a thermodynamic relationship and further developed a thermodynamic model based on density functional theory (DFT) calculations to explain the observed non-equimolar compositions of the boride and carbide phases quantitatively [32]. This model can be refined and extended to guide the design of other dual-phase HECs.

Such dual-phase HECs offer additional opportunities of microstructure engineering to improve the mechanical and other properties.

## 4. FROM HIGH-ENTROPY CERAMICS TO COMPOSITIONALLY COMPLEX CERAMICS

In 2020, we originally proposed to broaden "high-entropy ceramics (HECs)" to "compositionally complex ceramics (CCCs)" (Fig. 3), where non-equimolar and long- and short-range ordering, which reduce the configurational entropies, can provide additional dimensions or opportunities to tailor and improve the properties to outperform their equimolar, disordered, higher-entropy counterparts.

Using $(Hf_{1/3}Zr_{1/3}Ce_{1/3})_{1-x}(Y_{1/2}Yb_{1/2})_x O_{2-\delta}$ as an example, we demonstrated that a non-equimolar, medium-entropy composition holds a higher module ($E$) to thermal conductivity ($k$) ratio than its equimolar high-entropy counterpart (Fig. 3) [8]. This specific observation in YSZ-like quinary non-equimolar compositionally complex fluorite oxides can be understood from the clustering of oxygen vacancies.

This proposed extension from HECs to CCCs in ceramics is akin to the HEAs to CCAs expansion in metals, but the ceramic counterparts offer more complexities and opportunities with aliovalent doping, anion vacancies, and long- and short-range cation (and anion) ordering.

In addition, the spontaneously formed long- and short-range ordering (*e.g.*, cation ordering in $(V_{1/5}Cr_{1/5}Nb_{1/5}Ta_{1/5}W_{1/5})_5Si_3$, which stabilizing a new unusual phase [26] and the short-range weberite-type orders in high-entropy rare earth niobates that are in the disordered defect fluorite structure for the long-range order, which contributes to their ultralow thermal conductivities [30]) and dual-phase (or multi-phase) thermodynamic equilibria will reduce configurational entropies, driven by the thermodynamics.

Here, a key message is that the best properties and optimal performance often occur at non-equimolar compositions or with some long- and short-range orders. Maximizing the configurational entropy is not always necessary or desirable.

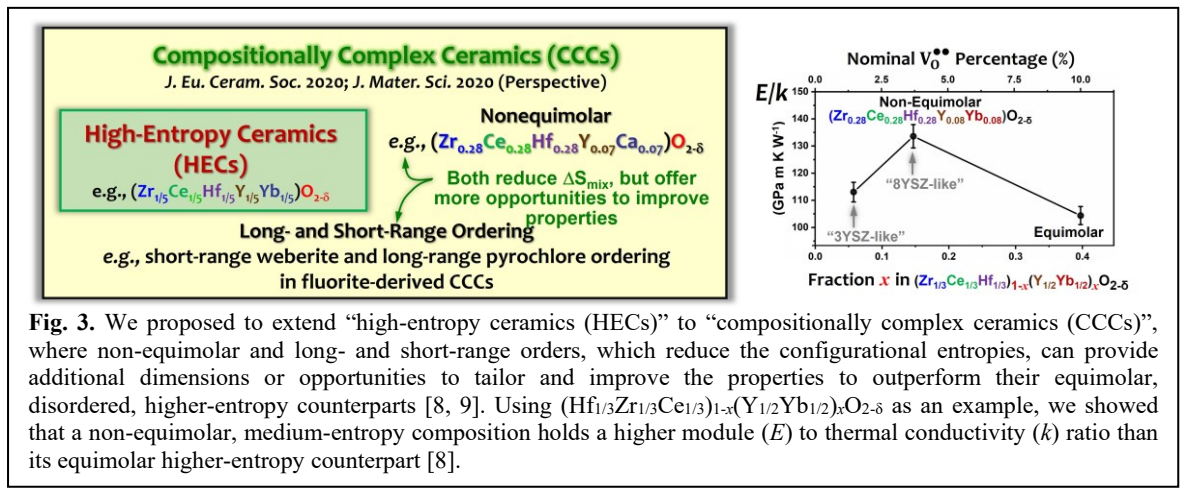

**Fig. 3.** We proposed to extend "high-entropy ceramics (HECs)" to "compositionally complex ceramics (CCCs)", where non-equimolar and long- and short-range orders, which reduce the configurational entropies, can provide additional dimensions or opportunities to tailor and improve the properties to outperform their equimolar, disordered, higher-entropy counterparts [8, 9]. Using $(Hf_{1/3}Zr_{1/3}Ce_{1/3})_{1-x}(Y_{1/2}Yb_{1/2})_x O_{2-\delta}$ as an example, we showed that a non-equimolar, medium-entropy composition holds a higher module ($E$) to thermal conductivity ($k$) ratio than its equimolar higher-entropy counterpart [8].

## 5. LONG- AND SHORT-RANGE ORDERING AND ULTRAHIGH-ENTROPY CERAMICS

A series of our recent studies investigated long- and short-range orders and order-disorder transitions (ODTs) in compositionally complex fluorite-based oxides (CCFBOs).

In 2018, we first reported the YSZ-like high-entropy fluorite oxides [27]. Subsequently, we fabricated a series of medium- and high-entropy ordered pyrochlore oxides and showed that size disorder (instead of the configurational entropy of mixing) controls their reduced thermal conductivities [28]. We further explored non-equimolar CCFBOs and demonstrated that they can outperform their equimolar, higher-entropy counterparts (Fig. 3) [8].

We further discovered composition-induced [10] and redox-induced [11] pyrochlore-fluorite ODTs in 10- and 11-compomnent CCFBOs. We showed the short-range weberite ordering causes ultralow thermal conductivity [30]. These provide additional opportunities to tailor and improve the thermal and mechanical properties.

Notably, we reported ultrahigh-entropy weberite and fergusonite phases, unusual stability of ultrahigh-entropy pyrochlore phases, and an unexpectedly abrupt pyrochlore-weberite transition in a 21-component system of CCFBOs [12].

On-going and unpublished results on the phase equilibria and transformations in CCFBOs will also be discussed at the conference.

This series of discoveries suggest great opportunities to use long- and short-range orders in these non-equimolar CCFBOs, and other HECs/CCCs in general, to tailor and improve their properties.

## 6. SUMMARY AND OUTLOOK

In comparison with metallic HEAs, HECs possess diversifying chemistries, crystal structures, and bonding characters. Combined with the vast compositional space, HECs bring substantial possibilities and opportunities as a new class of materials.

To date, the majority of studies of HECs focus on quinary (and occasionally 4- or 6-component) equimolar compositions. While HECs are still in their infancy stage, we proposed to broaden HECs to CCCs, as well as dual-phase and multi-phase HECs/CCCs, where non-equimolar compositional designs, long- and short-range orders, and microstructural engineering can bring additional and almost limitless opportunities to tailor and improve properties. Let us embrace the complexity!

Other important areas of studies include the synthesis and processing science and the fundamental interfacial science of the emerging classes of HECs and CCCs, which are substantially more complex than those for conventional ceramics due to the compositional complexity. However, the processing and interfacial sciences are of critical importance to enable the full potential of these emerging HECs and CCCs.

Notably, our recent on-going research also suggests that non-equimolar compositional designs are essential for developing two new classes of high-entropy and compositionally complex pervoskite oxides (HEPOs and CCPOs) as (1) active materials for solar thermochemical water splitting (for hydrogen generation) and (2) a new class of solid electrolytes for solid-state lithium-ion batteries. These studies again illustrate that maximizing the configurational entropy is not always desirable and optimal properties can often be attained at non-equimolar compositions considering various other materials design factors.

Finally, a most recent study by Yang et al. reported "high-entropy enhanced capacitive energy storage" in non-equimolar $(Bi_{3.25}La_{0.75})(Ti_{3-3x}Zr_xHf_xSn_x)O_{12}$ in the $Bi_2Ti_2O_7$-prototyped pyrochlore structure [33], which are rigorously not HECs but CCCs (based on the conventional definition [9]). This study of dielectric CCCs (in fact CCFBOs), along with our earlier study of YSZ-like CCFBOs for TBCs [8], further exemplifies the importance of exploring non-equimolar compositional designs (instead of simply maximizing the configurational entropy) in designing the future HECs and CCCs in general, and the high-entropy and compositionally complex dielectrics and ferroelectrics specifically.

ACKNOWLEDGEMENT:

The author acknowledges partial current supports from the U.S. National Science Foundation (NSF) – Grant No. DMR-2026193 in the Ceramics program for supporting research on compositionally complex fluorite-based oxides (CCFBOs) and Grant No. DMR-2011967 in the Materials Research Science and Engineering Center (MRSEC) program through the UC Irvine Center for Complex and Active Materials (CCAM) for supporting research on fundamental interfacial science of complex concentrated materials – and the U.S. Department of Energy (DOE), Office of Energy Efficiency and Renewable Energy (EERE), under the Agreement Number DE-EE0008839 (managed by the Hydrogen and Fuel Cell Technologies Office in the Fiscal Year 2019 H2@SCALE program), for supporting research on HEPOs/CCPOs for solar thermochemical hydrogen generation (via water splitting). The author is also grateful for prior supports from an ONR MURI program the DOE EERE Solar Technology Office. The author thanks his students (particularly Joshua Gild, Mingde Qin, Andrew Wright, Dawei Zhang, Sashank Shivakumar, and Naixie Zhou who directly contributed to our work on HECs and CCCs) and many collaborators.

REFERENCES:
[1] J. W. Yeh, S. K. Chen, S. J. Lin, J. Y. Gan, T. S. Chin, T. T. Shun, C. H. Tsau, and S. Y. Chang, *Advanced Engineering Materials* **6**, 299 (2004).
[2] B. Cantor, I. Chang, P. Knight, and A. Vincent, *Materials Science and Engineering*: A **375**, 213 (2004).
[3] D. B. Miracle and O. N. Senkov, *Acta Materialia* 122, 448 (2017).
[4] C. M. Rost, E. Sachet, T. Borman, A. Moballegh, E. C. Dickey, D. Hou, J. L. Jones, S. Curtarolo, and J.-P. Maria, *Nature Communications* **6**, 8485 (2015).
[5] J. Gild, Y. Zhang, T. Harrington, S. Jiang, T. Hu, M.

C. Quinn, W. M. Mellor, N. Zhou, K. Vecchio, and J. Luo, *Scientific Reports* **6**, 37946 (2016).
[6] S. Jiang, T. Hu, J. Gild, N. Zhou, J. Nie, M. Qin, T. Harrington, K. Vecchio, and J. Luo, *Scripta Materialia* **142**, 116 (2018).
[7] M. Qin, J. Gild, C. Hu, H. Wang, M. S. B. Hoque, J. L. Braun, T. J. Harrington, P. E. Hopkins, K. S. Vecchio, and J. Luo, *Journal of European Ceramic Society* **40**, 5037 (2020).
[8] A. J. Wright, Q. Wang, C. Huang, A. Nieto, R. Chen, and J. Luo, *Journal of European Ceramic Society* **40**, 2120 (2020).
[9] A. J. Wright and J. Luo, *Journal of Materials Science* **55**, 9812 (2020).
[10] A. J. Wright, Q. Wang, C. Hu, Y.-T. Yeh, R. Chen, and J. Luo, *Acta Materialia* **211**, 116858 (2021).
[11] D. Zhang, Y. Chen, T. Feng, D. Yu, K. An, R. Chen, and J. Luo, *Scripta Mater.* **215**, 114699 (2022).
[12] M. Qin, H. Vega, D. Zhang, S. Adapa, A. J. Wright, R. Chen, and J. Luo, *Journal of Advanced Ceramics* **11**, 641 (2022).
[13] L. Feng, W. G. Fahrenholtz, and D. W. Brenner, *Annual Review of Materials Research* **51**, 165 (2021).
[14] E. Castle, T. Csanádi, S. Grasso, J. Dusza, and M. Reece, *Scientific Reports* **8**, 8609 (2018).
[15] X. Yan, L. Constantin, Y. Lu, J.-F. Silvain, M. Nastasi, and B. Cui, *Journal of American Ceramic Society* **101**, 4486 (2018).
[16] T. J. Harrington, J. Gild, P. Sarker, C. Toher, C. M. Rost, O. F. Dippo, C. McElfresh, K. Kaufmann, E. Marin, L. Borowski, P. E. Hopkins, J. Luo, S. Curtarolo, D. W. Brenner, and K. S. Vecchio, *Acta Materialia* **166**, 271 (2019).
[17] J. Gild, A. Wright, K. Quiambao-Tomko, M. Qin, J. A. Tomko, M. Shafkat bin Hoque, J. L. Braun, B. Bloomfield, D. Martinez, T. Harrington, K. Vecchio, P. E. Hopkins, and J. Luo, *Ceram. Int.* **46**, 6906 (2020).
[18] J. Gild, K. Kaufmann, K. Vecchio, and J. Luo, *Scripta Materialia* **170**, 106 (2019).
[19] M. Qin, J. Gild, H. Wang, T. Harrington, K. S. Vecchio, and J. Luo, *Journal of European Ceramic Society* **40**, 4348 (2020).
[20] M. Qin, S. Shivakumar, T. Lei, J. Gild, E. C. Hessong, H. Wang, K. S. Vecchio, T. J. Rupert, and J. Luo, *Journal of European Ceramic Society* **42**, 5164 (2022).
[21] M. Qin, Q. Yan, H. Wang, C. Hu, K. S. Vecchio, and J. Luo, *Scripta Mater.* **189**, 101 (2020).
[22] M. Qin, Q. Yan, Y. Liu, and J. Luo, *Journal of Advanced Ceramics* **10**, 166 (2021).
[23] M. Qin, Q. Yan, H. Wang, K. S. Vecchio, and J. Luo, *Journal of European Ceramic Society* **41**, 2968 (2021).
[24] M. Qin, Q. Yan, Y. Liu, H. Wang, C. Wang, T. Lei, K. S. Vecchio, H. L. Xin, T. J. Rupert, and J. Luo, *Journal of European Ceramic Society* **41**, 5775 (2021).
[25] J. Gild, J. Braun, K. Kaufmann, E. Marin, T. Harrington, P. Hopkins, K. Vecchio, and J. Luo, *Journal of Materiomics* **5**, 337 (2019).
[26] S. Shivakumar, M. Qin, D. Zhang, C. Hu, Q. Yan, and J. Luo, *Scripta Materialia* **212**, 114557 (2022).
[27] J. Gild, M. Samiee, J. L. Braun, T. Harrington, H. Vega, P. E. Hopkins, K. Vecchio, and J. Luo, *Journal of European Ceramic Society* **38**, 3578 (2018).
[28] A. J. Wright, Q. Wang, S.-T. Ko, K. M. Chung, R. Chen, and J. Luo, *Scripta Mater.* **181**, 76 (2020).
[29] A. J. Wright, C. Huang, M. J. Walock, A. Ghoshal, M. Murugan, and J. Luo, *Journal of American Ceramic Society* **104**, 448 (2021).
[30] A. J. Wright, Q. Wang, Y.-T. Yeh, D. Zhang, M. Everett, J. Neuefeind, R. Chen, and J. Luo, *Acta Mater.* **235**, 118056 (2022).
[31] N. Zhou, S. Jiang, T. Huang, M. Qin, T. Hu, and J. Luo, *Science Bulletin* **64**, 856 (2019).
[32] M. Qin, J. Gild, C. Hu, H. Wang, M. S. B. Hoque, J. L. Braun, T. J. Harrington, P. E. Hopkins, K. S. Vecchio, and J. Luo, *Journal of European Ceramic Society* **40**, 5037 (2020).
[33] B. Yang, Y. Zhang, H. Pan, W. Si, Q. Zhang, Z. Shen, Y. Yu, S. Lan, F. Meng, and Y. Liu, *Nature Materials*, in press (2022).